\documentclass[reprint,superscriptaddress,prx]{revtex4-2}

\usepackage{xcolor}
\usepackage{subcaption}
\usepackage{mathtools}
\usepackage{overpic}
\usepackage{mathrsfs}
\usepackage{psfrag}
\usepackage{amsfonts,bbm}
\usepackage[bbgreekl]{mathbbol}
\usepackage{braket}
\usepackage{caption}
\usepackage{import}
\usepackage{transparent}

\usepackage{hyperref}

\newcommand{\ve}[1]{\boldsymbol{#1}} 

\newcommand{\obsc}[1]{\textcolor{blue}{#1}}

\begin{document}

\title{Steer’n Roll: A Stereoscopic Flow-Sensing Strategy for Planktonic Prey Detection and Capture}

\date{\today}

\author{Tommaso Redaelli}
\affiliation{Aix Marseille Univ, CNRS, Centrale Med, IRPHE, Marseille, France}

\author{Eva Kanso}
\email{kanso@usc.edu}
\affiliation{Department of Aerospace and Mechanical Engineering, University of Southern California, Los Angeles, CA, USA}

\author{Christophe Eloy}
\email{christophe.eloy@centrale-med.fr}
\affiliation{Aix Marseille Univ, CNRS, Centrale Med, IRPHE, Marseille, France}

\begin{abstract}
Planktonic organisms such as copepods sense swimming prey and sinking food particles through the hydrodynamic disturbances they generate. However, because these flow fields are often highly symmetric, they provide little directional information, making accurate localization of the source challenging. Here, we introduce the ``steer’n roll'' sensing and response strategy. This strategy combines stereoscopic flow sensing and a roll motion. Stereoscopic sensing allows plankton to disambiguate flow signals by integrating two spatially separated flow measurements, while a roll about the swimming axis enhances exploration of the three-dimensional space. We show that steer’n roll is efficient, achieving a 100\% success rate, versatile across signal type, and robust to flow sensing noise, orientational diffusion, and turbulence. Together, these findings identify a biologically plausible mechanism for prey detection and capture via flow sensing, and offer testable insights into the sensory-motor strategies of planktonic organisms.
\end{abstract}

\maketitle

\section{Introduction}
	

	Mesoplankton are a group of millimetric organisms (ranging from 0.2 to 20 mm in size) that form a substantial component of the ocean's biomass and play a central role in structuring marine food webs and regulating the global carbon cycle~\cite{steinberg2017,hays2005}. 
	Among them, copepods dominate both in abundance and biomass \cite{Longhurst1985,Humes1994,Verity1996} and inhabit virtually all 
	marine environments, from shallow coastal to deep waters \cite{Hardy1970}. These crustaceans are often omnivorous; they feed on different types of prey~\cite{Kiorboe1995,Kiorboe2013}, from non-motile diatoms, marine snow, or fecal pellets to motile organisms, such as ciliates, fish larvae, or nauplii of other copepod species \cite{Calbet2007}.
	
	Despite their ecological importance, copepods are visually impaired. Although they possess an eye spot that detects light, they are unable to form images \cite{Stearns1984}. 
	Instead, they depend on hydrodynamic and chemical sensing to locate prey and avoid predators \cite{Visser2001,strickler2007,Jiang2008,kiorboe2010danger,Tuttle2019, LegierVisser1986}. 
	To detect flow disturbances and chemical gradients in a turbulent ocean, copepods use two types of hair-like sensors that are mainly found on their antennules (Fig.\ref{fig:fig_1_scheme}A): setae, sensitive to flow-induced loads \cite{Strickler1973b,Derby2014,Yen2015}, and aesthetascs, chemical sensors sensitive to different acids and proteins \cite{Lenz1996}. 
	The diversity of setal lengths suggests specialization across different ranges of flow velocity \cite{Fields2002}, endowing copepods with exceptional flow sensitivity and frequency response unmatched across crustaceans~\cite{Weatherby2000}. 
	
	Consistent with these sensory capabilities, laboratory experiments have demonstrated that copepods respond to a wide range of hydrodynamic disturbances, including the motion of small cylinders \cite{Buskey2002}, swimming ciliates \cite{Jiang2008}, imposed shear flows~\cite{Kiorboe1999}, and background turbulence~\cite{Webster2015,Michalec2015,Michalec2017}. At the same time, turbulence has been shown to impair copepods escape responses, highlighting the challenges of extracting reliable information in complex flow environments~\cite{Gilbert2005,Robinson2007,Waggett2007}. 
	
	How do copepods detect and localize potential sources of food or threats using hydrodynamic signals? 
	Being planktonic and transported by water currents, copepods cannot sense the mean flow velocity, they can only sense local gradients of the surrounding flow \cite{Kiorboe1995,Estrada1997,Pepper2015}.
	Determining the position and orientation of a flow source from such local measurements poses a complex inverse problem that cannot be solved with a single measurement of the flow gradient~\cite{Colvert2017a, Takagi2020,Borra2022,takagi2017}. 
	Indeed, the flow generated by a swimming or sinking prey is both axially and centrally symmetric, making it impossible for a predator to disambiguate left from right and front from back by relying on a single measurement of the local flow gradient.

Intriguingly, many microorganisms, including copepods~\cite{Niimoto2020} and dinoflagellates~\cite{Fenchel2001}, exhibit a rotation around their swimming axis, producing helical trajectories. 
In algae, this rolling motion has been shown to enable phototaxis using only a single light sensor by allowing sequential sampling of the three-dimensional space~\cite{wan2021origins, Cortese2021, Leptos2023}.
More generally,  rolling motions allow organisms with planar sensing to augment their sensory dimensionality to the full three-dimensional space. We propose that this behavior represents an evolved strategy, increasing the three-dimensional perception of signals. 
	
	In this article, we propose a versatile and robust ``steer'n roll'' sensing-response strategy that resolves intrinsic ambiguities in hydrodynamic signals. This strategy relies on stereoscopic flow sensing achieved through two local measurements of the flow gradient, which together allow a predator to determine the funneling direction of the flow signal. We will show that steering along this direction, combined with rolling about the swimming direction, enables reliable localization of prey. 
	To assess the versatility of steer'n roll, we quantify capture success and demonstrate that performance is independent of the signal type. We then examine the robustness of the strategy to sensory noise, orientational diffusion, and background turbulence.
	Finally, we discuss the ecological implications of our findings for planktonic organisms.


	

	
	
	\section{Hydrodynamic Signatures of Planktonic Prey}
	\subsection{Mathematical formulation}
We formulate prey detection and localization as a mathematical problem of hydrodynamic sensing in viscous fluid. 
We begin by characterizing the flow fields generated by microscale prey and the constraints faced by predators in interpreting these signals. We then identify the geometric features of the flow that can be exploited to enable successful prey localization.
    
The flow generated by a prey of size $a $ (1--100~$\mu$m) moving at a speed $U $ (1--100~$\mu$m s$^{-1}$) is characterized by a low Reynolds number, $\mathrm{Re} = \rho a U /\mu = 10^{-6}$--$10^{-2}$, where $\rho = 1000\;$kg\;m$^{-3}$ and $\mu \approx 10^{-3}\;$Pa$\;$s are the density and viscosity of water. In this regime, fluid dynamics is governed by viscous forces and the fluid velocity field $\ve v$ generated by the prey is described by the  Stokes equations \cite{happelbrenner1981}. 

\begin{figure}[ht!]
    \centering
   \def\svgwidth{0.5\textwidth}
		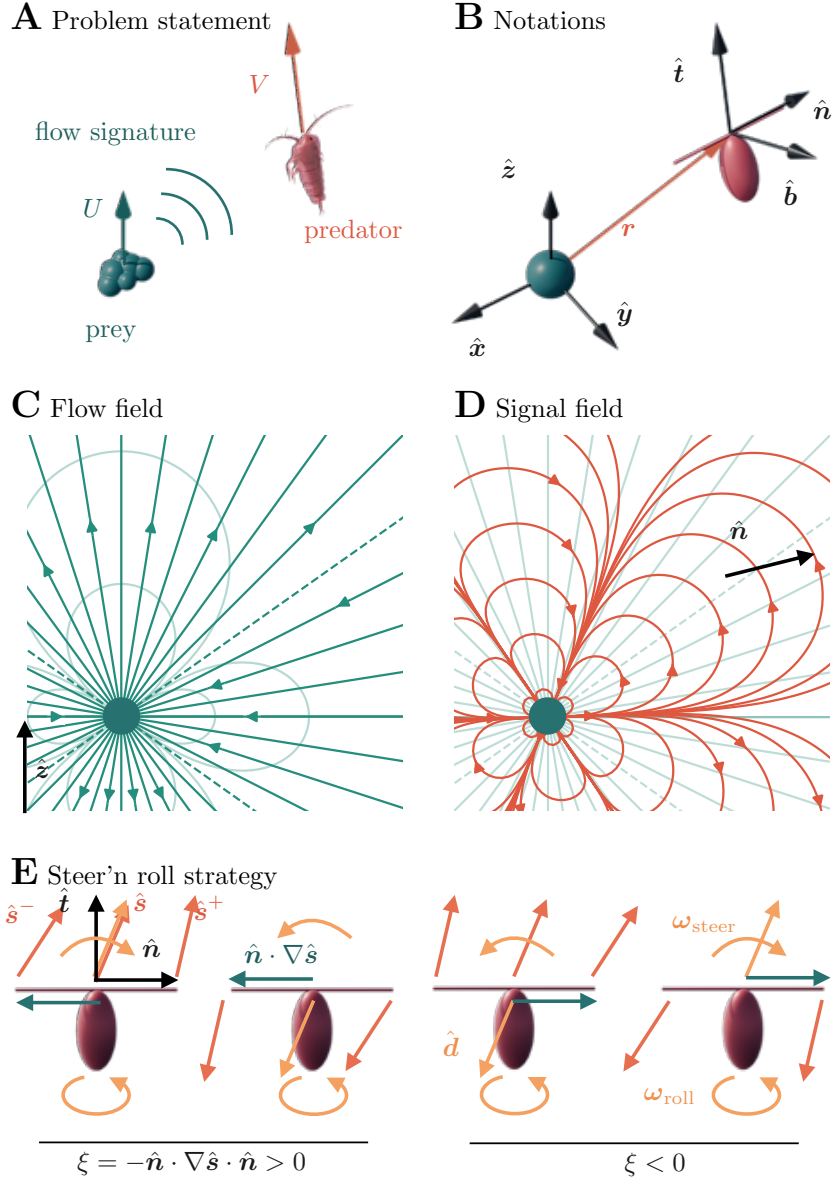
    \caption[Scheme]{\textbf{Overview of the predator--prey sensing problem and the steer'n roll strategy.}
    (\textbf{A}) Schematic view of the predator-prey problem. The prey, swimming (or sinking) at speed $U$, generates a fluid disturbance that the predator detects through mechanosensors. The predator speed swims at constant speed $V$. 
    (\textbf{B}) Reference frames: a prey-fixed Cartesian frame $(\boldsymbol{\hat x}, \boldsymbol{\hat y}, \boldsymbol{\hat z})$ and and a predator-fixed body frame $(\boldsymbol{\hat n}, \boldsymbol{\hat b}, \boldsymbol{\hat t})$. The vector $\boldsymbol{r}$ denotes the position of the predator in the prey's reference frame. 
    (\textbf{C}) Streamlines (dark green) and velocity-magnitude contours (light green) of the stresslet flow field (~\eqref{eq:Stresslet}) generated by a swimming prey .
    (\textbf{D}) Signal field \(\mathbf{S}\) (red) (~\eqref{eq:dimensional_signal}) measured by a predator with sensors aligned along \(\boldsymbol{\hat n}\) (black). Prey-generated flow streamlines are shown in light green for reference.
    (\textbf{E}) Steer'n roll strategy: red arrows indicate local signal directions \(\boldsymbol{\hat s}\); orange arrow \(\boldsymbol{\hat d}\) denotes the inferred prey direction obtained from stereoscopic sensing (~\eqref{eq:triangulation_d}). The angular velocity \(\boldsymbol{\omega}_{\mathrm{steer}}\) steers the predator toward \(\boldsymbol{\hat d}\), while \(\boldsymbol{\omega}_{\mathrm{roll}}\) induces active rotation about the swimming direction \(\boldsymbol{\hat t}\). The scalar \(\xi\) (\eqref{eq:triangulation_sign}) determines the funneling direction of the signal field.}
    \label{fig:fig_1_scheme}
\end{figure}

Let $(\boldsymbol{\hat x}, \boldsymbol{\hat y}, \boldsymbol{\hat z})$ denote the frame of reference centered on the prey (Fig.~\ref{fig:fig_1_scheme}B), with axis $\boldsymbol{\hat z}$ aligned to the prey moving direction (either the swimming or sinking direction) which is generally a symmetry axis of the prey-generated flow. Let $\ve r$ be the position of the predator in the $(\boldsymbol{\hat x}, \boldsymbol{\hat y}, \boldsymbol{\hat z})$ reference frame. In the following, we will also use a set of cylindrical coordinates with associated unit vectors  $(\boldsymbol{\hat \rho}, \boldsymbol{\hat \varphi}, \boldsymbol{\hat z})$ sharing the same $\boldsymbol{\hat z}$-axis. 
    
   We define a body-fixed frame  $(\boldsymbol{\hat t}, \boldsymbol{\hat n}, \boldsymbol{\hat b})$ attached to the predator, where $\boldsymbol{\hat t}$ points in the predator's swimming direction, the predator's sensors are located along $\boldsymbol{\hat n}$, and $\boldsymbol{\hat b}$ denotes the binormal (Fig.~\ref{fig:fig_1_scheme}B). The predator is modeled as a point particle swimming at constant speed $V$.
    We assume that the predator's morphological features and motion do not perturb the flow field generated by the prey. Although both predator and prey may be advected by background flows, we fix the prey to simplify the analysis and enable a tractable derivation of the sensing strategy. This assumption holds true when $V \gg U$.

	%
%
	
\subsection{Prey flow signature}
We aim for a general description of the flow generated by prey that does not depend on  individual-specific 
details, in order to capture the diversity of plankton.
Because the prey flow field $\ve v$ is a solution of the Stokes equations, by linearity, it can be expressed as a sum of essential singularities 
	\begin{equation}
		\begin{split}
			\ve v(\ve r) & = \ve v_1(\ve r) + \ve v_2(\ve r) + \ldots, \\
			& =  \dfrac{1}{8\pi\mu} \left( \boldsymbol{F}\cdot \mathbb{G} + \mathbb{D}\cdot \nabla \mathbb{G} + \ldots \right),
			\label{eq:expansion}
		\end{split}
	\end{equation}
	using the Green-Oseen tensor $\mathbb{G}={\mathbb{I}}/{r} + ({\ve r \otimes\ve r})/{r^3}$ and its higher-order derivatives. Here, $\mathbb{I}$ is the identity tensor, $\otimes$ the tensor product, and $r = \| \ve r\|$~\cite{happelbrenner1981, pozrikidis1992, dhont1996, chwang1975}.

	Sinking preys, such as diatoms and marine snow particles, apply on the fluid, at leading order, a force monopole $\boldsymbol{F} = -F \delta(\ve r)\boldsymbol{\hat z}$ of strength $F$ equal to  their apparent weight, 
    %
    $\delta(\ve r)$ being the Dirac-delta function; the corresponding non-zero term in~\eqref{eq:expansion} is the stokeslet solution 
    \begin{equation}
        \ve v_{\rm 1}(\ve r)=\boldsymbol{F} \cdot\mathbb{G} /8\pi\mu\:,
\label{eq:Stokeslet}
    \end{equation}
     which decays as $1/r$~\cite{happelbrenner1981}.
    
	Swimming preys exert, at leading order, a force dipole $\mathbb{D} = \pm F a \delta(\ve r)\boldsymbol{\hat z}\otimes\boldsymbol{\hat z}$ on the fluid~\cite{lauga2009}. The dipole results from a force balance between swimming thrust $F\boldsymbol{\hat{z}}$ and viscous drag $-F\boldsymbol{\hat{z}}$ applied at points separated by a distance $a$ along $\boldsymbol{\hat{z}}$. The sign of the dipole depends on the type of swimmer:  positive for \textit{pullers} such as the algae \textit{Chlamydomonas} and negative for \textit{pushers} such as spermatozoa~\cite{lauga2009,dresher2010}. The associated flow is a  stresslet of the form (Fig.~\ref{fig:fig_1_scheme}C)
    \begin{equation}
        \ve v_{\rm 2}(\ve r)= \mathbb{D} \cdot \nabla \mathbb{G} /{8\pi\mu}\:,
    \label{eq:Stresslet}
    \end{equation}
    which decays as $1/r^2$~\cite{lauga2009}.
	
	In general, any projection of the $n^{th}$-order derivatives 
	$\nabla^n\mathbb{G}$ provides a solution of the Stokes equations of order $n+1$.
	In what follows, to limit mathematical complexity, we retain only the first two terms of the multipole expansion in~\eqref{eq:expansion} and adopt a simplified description of the flow field generated by the prey. This expansion, however, applies more broadly: linear combinations of higher-order terms can capture the full hydrodynamic complexity of any prey geometry and motion~\cite{kim2005}.

\subsection{Signal detected by predator}
Inspired by copepod biology, we consider that the predator is equipped with flow sensors positioned along its antennae, parallel to the vector $\boldsymbol{\hat{n}}$. Assuming that the predator is drifting with velocity $\ve v(\ve r)$, a flow sensor located at $\ve r_s = \ve r + \ell \boldsymbol{\hat n}$ measures the relative velocity signal (Fig.~\ref{fig:fig_1_scheme}D)
	\begin{equation}
		\ve S(\ve r_s) = \ve v(\boldsymbol{r}_s) - \ve v(\ve r)
		\approx  \ell \,\boldsymbol{\hat n} \cdot \nabla  \ve v (\ve r_s).
        \label{eq:dimensional_signal}
	\end{equation}
	Here, we approximated the flow signal $\ve S$ by its leading-order expansion, assuming $\ell\ll r$~\cite{Colvert2017}.  
    
    By analogy with~\eqref{eq:expansion}, 
	the field $\ve S(\ve r)$ is itself a solution of the Stokes equation, and corresponds to a multipole of one order higher than that of the flow field $\ve v(\ve r)$. Consequently, the signal vector field $\ve S(\ve r)$ can be decomposed into fundamental Stokes singularities located at the prey position. 
	
	The flow fields $\ve v_1(\ve r)$ and $\ve v_2(\ve r)$ associated with sinking and swimming prey, respectively, are both axially symmetric under rotations around  $\boldsymbol{\hat z}$ and centrally symmetric. The parity of this central symmetry depends on the order of the Stokes singularity: , 
	$\ve v_1(\ve r)=\ve v_1(-\ve r)$ is even, whereas $\ve v_2(\ve r)=-\ve v_2(\ve -r)$ is odd. The corresponding signal field $\ve S(\ve r)$ inherits the opposite parity:  odd when the velocity field is even, and vice versa. These symmetry properties make it impossible for a predator to disambiguate the prey location from a single measurement~\cite{takagi2017}. 

\section{Stereoscopic Flow Sensing}
To overcome the intrinsic ambiguities of flow signals, we introduce ``steer'n roll,'' a flow sensing and response strategy in which a predator combines stereoscopic flow sensing and active reorientation to successfully detect and capture its prey.
    
\subsection{Stereoscopic flow sensing reveals prey direction} 
The steer'n roll strategy resolves the ambiguities imposed by the symmetries of the flow and signal fields by exploiting two simultaneous flow measurements. This stereoscopic flow sensing is inspired by the binaural cues for sound localization used by mammals \cite{Grothe2010}.  

Specifically, we consider a predator equipped with two flow sensors, one on each antenna, located at $\boldsymbol{r}_s^\pm =\boldsymbol{r} \pm \ell \boldsymbol{\hat n}$, where $\ell$ denotes the antennule length and the superscript $\pm$ distinguishes the left and right antennules. 
With these two sensors, the predator has access to two flow signals 
	\begin{equation}
		\ve S^\pm = \pm \ell \,\boldsymbol{\hat n} \cdot \nabla \ve v(\boldsymbol{r}_s^\pm)
        \approx \ve S (\ve r) \pm \ell \boldsymbol{\hat n} \cdot \nabla \ve S (\ve r).
		\label{eq:stereosignal}
	\end{equation}
	These signals define two directions, which we represent as unit vectors (Fig.~\ref{fig:fig_1_scheme}E)
	\begin{equation}
		\boldsymbol{\hat s}^\pm = \frac{\ve S^\pm }{ \| \ve S^\pm \|}\:.
    \label{eq:signals}
	\end{equation}
     The unit vectors $\boldsymbol{\hat s}^\pm$ are aligned with the field lines of $\ve S$. Because $\ve S$ is a singular solution of the Stokes equations, its field lines converge to the location of the singularity, which corresponds to the prey's location. 
     
Stereoscopic flow sensing proceeds in two steps. First, we reconstruct a three-dimensional direction $\boldsymbol{\hat d}$ aligned with the funneling direction of the signal streamlines (Fig.~\ref{fig:fig_1_scheme}E). Geometrically,  $\boldsymbol{\hat d}$ points towards the location of the shortest distance between the two lines passing by the points $\boldsymbol{r}_s^\pm$ with unit vectors $\boldsymbol{\hat s}^\pm$  (see \textit{Supporting Information} Sec.~\ref{sec:AppTriang})\obsc{(see SI)}. Namely, 
\begin{equation}\label{eq:triangulation_d}
		\boldsymbol{\hat d}  =  \mathrm{sign}(\xi) \boldsymbol{\hat s}\:, 
\end{equation}
with $\boldsymbol{\hat s} = (\boldsymbol{\hat{s}}^+ + \boldsymbol{\hat{s}}^-)/\| \boldsymbol{\hat{s}}^+ + \boldsymbol{\hat{s}}^-\|\approx {\ve S}/{ \| \ve S \|}  $ and
		\begin{equation}
        \label{eq:triangulation_sign}
    \xi = - \frac{1}{2\ell}
    \boldsymbol{\hat n} \cdot (\boldsymbol{\hat{s}}^+ - \boldsymbol{\hat{s}}^-)
    \approx - \boldsymbol{\hat n} \cdot \nabla \boldsymbol{\hat{s}}
    \cdot \boldsymbol{\hat n}\:. 
		\end{equation}
        
Second, the predator steers in the direction $\boldsymbol{\hat d}$. 
To this end, we prescribe an angular velocity 
	\begin{equation}
		\boldsymbol{\omega}_{\rm steer} = {\omega}_{\rm steer} \frac{\boldsymbol{\hat t} \times \boldsymbol{\hat d}}{\| \boldsymbol{\hat t} \times \boldsymbol{\hat d}\|},
		\label{eq:omega_steer}
	\end{equation}
which rotates with angular speed ${\omega}_{\rm steer}$ toward $\boldsymbol{\hat d}$.
The subtlety here is that $\boldsymbol{\omega}_{\rm steer}$ depends on the orientation of the predator through $\boldsymbol{\hat t}$  because $\boldsymbol{\hat d}$ depends on $\boldsymbol{\hat s}$, which depends on $\boldsymbol{\hat n}$.

	\section{Steer'n Roll and Resulting Predator Trajectories}

   \subsection{Steer'n roll strategy} 
   We now formalize the steer’n roll strategy as a dynamical system describing the coupled translational and rotational motion of the predator. The strategy consists of three elementary actions: (i) steering toward the funneling direction of the hydrodynamic signal, (ii) rolling about the swimming axis, and (iii) swimming at constant speed.

In addition to the steering angular velocity \(\boldsymbol{\omega}_{\rm steer}\) defined in~\eqref{eq:omega_steer}, we introduce a rolling angular velocity $\boldsymbol{\omega}_{\rm roll}$ about the swimming axis, inspired by the rolling motion observed in marine organisms~\cite{Niimoto2020,Fenchel2001,wan2021origins, Cortese2021, Leptos2023},
\begin{equation}
	   \boldsymbol{\omega}_{\rm roll}  = \omega_{\rm roll} \, \boldsymbol{\hat t}\:.
       \label{eq:omega_roll}
	\end{equation}
	The predator’s motion obeys the dynamical system
    \begin{equation}
		\begin{split}
			\label{Eq:Dynamics}
			\frac{d \ve r}{dt}  = V\boldsymbol{\hat t},\qquad 
			\frac{d \boldsymbol{\hat e}}{dt}  =(\boldsymbol{\omega}_\mathrm{steer} + \boldsymbol{\omega}_\mathrm{roll})\times \boldsymbol{\hat e}, 
		\end{split}
	\end{equation}
	where  $\boldsymbol{\hat e}$ denotes any of the body-fixed unit vectors $\boldsymbol{\hat n}$, $\boldsymbol{\hat b}$, or $\boldsymbol{\hat t}$. This formulation makes explicit that steer’n roll combines directional guidance from flow sensing with active reorientation, rolling and forward swimming.

\begin{figure*}[t]
    \centering
    \includegraphics[width=\textwidth]{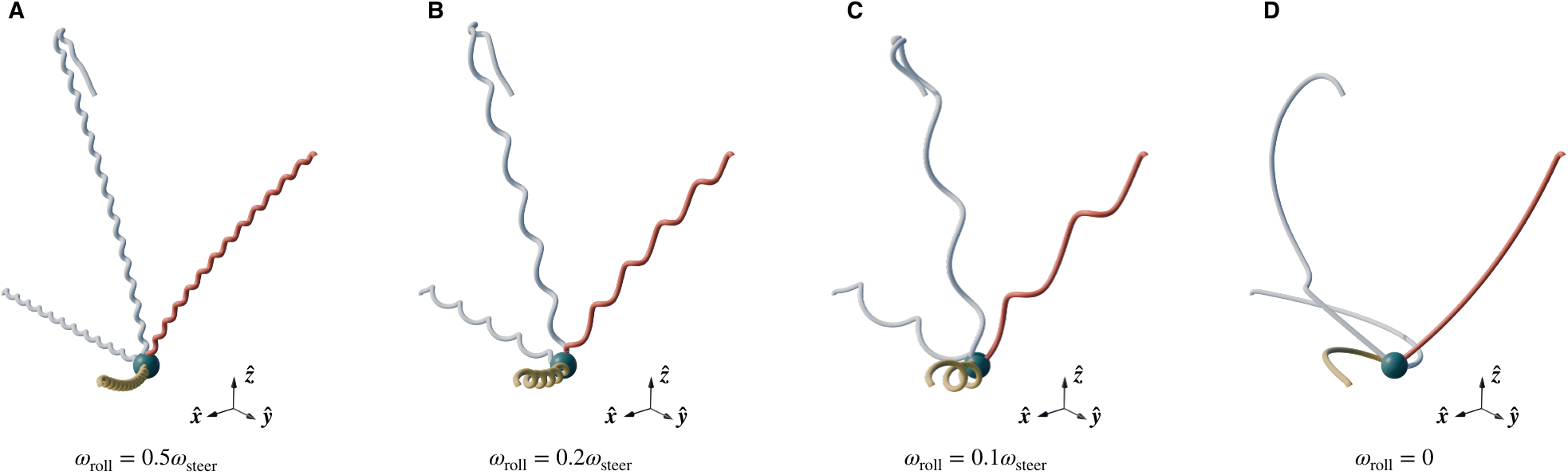}
    \caption[Scheme]{
\textbf{Predator trajectories.} 
Trajectories obtained by integrating \eqref{Eq:Dynamics} for a prey emitting a stresslet $\ve v_2(\ve r)$.
We simulate four different initial predator's positions, all with the same initial distance $r_0$ from the prey (green sphere) and same initial orientation.  
The predator speed is $V = 0.01r_0\omega_\mathrm{steer}$ and the prey is represented as a sphere of radius $a=0.04 r_0$.
Four values of the roll speed $\omega_{\rm roll}$ are used as labeled. 
The trajectories are helices with a radius decreasing with $\omega_{\rm roll}$, but similar pitch angle. For two representative trajectories of panel B (red and yellow), the dynamics of the predator's orientation $\boldsymbol{\hat{t}}$ is shown in Fig.~\ref{fig:fig_3_t_trajectories}A, and the dynamics of the predator's sensory antennae $\boldsymbol{\hat{n}}$ in Fig.~\ref{fig:Trajectories_vector_n} in the \textit{Supporting Information}.}
    \label{fig:fig_2_trajectories}
\end{figure*}

\subsection{Numerical integration}  
We solve~\eqref{Eq:Dynamics} numerically using a second-order Runge–Kutta integration scheme. Figure~\ref{fig:fig_2_trajectories} shows representative trajectories of a predator responding to a swimming prey modeled as a three-dimensional stresslet, $\ve v_2(\ve r)$. 
It shows that the predator is always successful regardless of its initial position $\ve r_0$ or orientation $\boldsymbol{\hat t}$. 

When the roll is non zero, trajectories converge towards helices of constant pitch angle and decreasing radius with increasing $V/\omega_\mathrm{roll}$. 
As discussed in the following, these helical trajectories can be interpreted physically by examining the fixed points of \eqref{Eq:Dynamics}.

\subsection{The role of roll} 
According to the steering motion in \eqref{eq:omega_steer}, the predator can in principle execute both yaw (rotation around $\boldsymbol{\hat b}$) and pitch (rotation around $\boldsymbol{\hat n}$). This is sufficient to solve the symmetry-induced ambiguities of Stokes flows, and the addition of roll does not significantly affect the success of the strategy in prey localization (Fig.~\ref{fig:fig_2_trajectories}). 

However, if motor capabilities do not allow for pitch but only for yaw,  $\boldsymbol{\omega}_{\rm steer}$ is projected onto $\boldsymbol{\hat b}$. In this case, motion is restricted to the initial $(\boldsymbol{\hat{n}},\boldsymbol{\hat{t}})$--plane, which in general does not contain the flow source (i.e., prey location) and prevents exploration of the full three-dimensional space. Introducing a nonzero roll motion $\boldsymbol{\omega}_{\rm roll} \neq 0$, restores access to the full three-dimensional environment. A combined steer-and-roll strategy therefore enables effective navigation and prey localization even under restricted motor capabilities (see \textit{Supporting Information} Fig.~\ref{fig:RollScalar}).

\section{Convergence to the Prey}
		
\subsection{Fixed points of the orientation dynamics}
To better understand the origin of the helical trajectories in Fig.~\ref{fig:fig_2_trajectories}, we calculate the fixed points of the orientation dynamics. We place ourselves in the far-field limit ($r \gg V / \omega_\mathrm{steer}$), and asymptotically small roll ($\omega_\mathrm{roll} \ll \omega_\mathrm{steer}$). 
We thus consider $V=0$ and $\omega_\mathrm{roll} = 0$ in \eqref{Eq:Dynamics}, and look for the fixed points $\{\boldsymbol{\hat t^*}, \boldsymbol{\hat n^*}, \boldsymbol{\hat b^*}\}$ of 
the orientation dynamics satisfying 
	\begin{equation}\label{eq:fixedpoint}
		\boldsymbol{\hat t}^* \times \boldsymbol{\hat s} \left( \boldsymbol{\hat n}^* \right)
        = \boldsymbol{\hat t}^* \times \left(
            \boldsymbol{\hat n}^* \cdot \nabla \ve v
        \right)
		= 0,
	\end{equation}
    which expresses the condition that the swimming direction $\boldsymbol{\hat t}$ aligns with the local direction of the signal field $\boldsymbol{\hat s}$.
    
To solve~\eqref{eq:fixedpoint}, we consider a generic gradient $\nabla \ve v$ of the prey-generated flow. This gradient has two useful properties: because the flow is incompressible, $\nabla \boldsymbol{v}$ is traceless, and because the flow is axisymmetric,  the azimuthal direction $\boldsymbol{\hat \varphi}$ is an eigenvector. 
In other words, when expressed using the eigenbasis $(\boldsymbol{\hat e_1}, \boldsymbol{\hat e_2}, \boldsymbol{\hat e_3})$ associated with its symmetric part (with $\boldsymbol{\hat e_3} = \boldsymbol{\hat \varphi}$), the gradient $\nabla \ve v$ takes the form
	\begin{equation}\label{eq:gradient}
		\nabla \boldsymbol{v} = \lambda_1
		\begin{bmatrix}
			1 		& \Omega 	& 0 \\
			-\Omega & \tilde{\lambda} 		& 0 \\
			0 		& 0 		& -1-\tilde{\lambda}
		\end{bmatrix}.
	\end{equation}
Here, $\lambda_1$ is the largest eigenvalue (in absolute value) of the symmetric part, $\tilde{\lambda} = \lambda_2 / \lambda_1$ is the ratio of the eigenvalues in the radial plane, and $\Omega$ is the relative rotation associated to the skew-symmetric part. 
Thus, the normalized velocity gradient (i.e., $\nabla \boldsymbol{v} /\lambda_1$) depends only on two parameters: $-1\le \tilde{\lambda}\le 1$ and $\Omega$.  
	
Substituting Eqs. (\ref{eq:stereosignal}) and (\ref{eq:gradient}) into~\eqref{eq:fixedpoint},  we obtain (see \textit{Supporting Information})
	\begin{subequations}		  \label{eq:fp}
		\begin{align}
            \frac{\tilde{\lambda}-1}{\tilde{\lambda}} t_2^{*2} + 
		      \frac{\tilde{\lambda}^2 + 2\tilde{\lambda} + \Omega^2}{\tilde{\lambda}^2 + \tilde{\lambda}} t_3^{*2} &= 1,\\
            (1-\tilde{\lambda}) n_2^{*2} + (2+\tilde{\lambda}) n_3^{*2} &= 1,
        \end{align}
	\end{subequations}
where $t_i^{*}$ $n_i^{*}$ are the components of $\boldsymbol{\hat{t}}^*$ and $\boldsymbol{\hat{n}}^*$ along $\boldsymbol{\hat e}_i$, with $i = 1,2,3$. For $\tilde{\lambda}<0$, a condition satisfied by all prey-generated flows considered here, each equation describes an elliptical cylinder with axis along $\boldsymbol{\hat e_1}$. Since $\sum_i (t_i^{*})^2=\sum_i (n_i^{*})^2=1$, the fixed points $\boldsymbol{\hat{t}}^*$ and $\boldsymbol{\hat{n}}^*$ lie at the intersection of these elliptical cylinders and the unit sphere. 

	\begin{figure*}[t]
		\centering
		\def\svgwidth{\textwidth}
		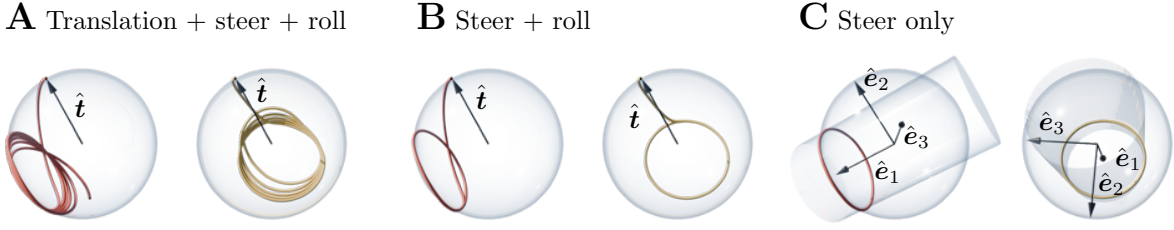
	\caption[t trajectories]{
    \textbf{Predator's orientation dynamics.}
(\textbf{A}) Evolution of \(\boldsymbol{\hat t}\) extracted from two of the trajectories shown in Fig.~\ref{fig:fig_2_trajectories}B (red and yellow). The dynamics converges to an imperfect limit cycle.
(\textbf{B}) Same as A, but in the absence of rotation (\(V=0\)), showing convergence to a perfect limit cycle.
(\textbf{C}) Theoretical limit cycle predicted for $V=0$ and $\omega_\mathrm{roll} = 0$, by the fixed-point analysis (see~\eqref{eq:fp}).}
	\label{fig:fig_3_t_trajectories}
	\end{figure*}	
    
\subsection{Limit cycles and helical trajectories}
We now compare these analytical predictions with the result of the simulations. For this purpose, the predator's orientation $\boldsymbol{\hat{t}}$ has been extracted from two of the trajectories shown in Fig.~\ref{fig:fig_2_trajectories}B. It shows that, after a transient, $\boldsymbol{\hat{t}}$ converge towards a limit cycle (Fig.~\ref{fig:fig_3_t_trajectories}A). However, this convergence is not perfect because the relative position of the predator with respect to the prey changes along the trajectory. By removing the translation from the dynamics ($V=0$), the vector $\boldsymbol{\hat{t}}$ converge towards a perfect limit cycle as shown in Fig.~\ref{fig:fig_3_t_trajectories}B. 

This limit cycle can be interpreted from the above analysis as the intersection of an elliptical cylinder given by Eq.~(\ref{eq:fp}a) with the unit sphere (Fig.~\ref{fig:fig_3_t_trajectories}C). In principle, this intersection defines two distinct limit cycles. 
One of the two cycles is selected when adding the condition $\xi>0$, with $\xi$ given by \eqref{eq:triangulation_d}: the predator chooses the direction  $\boldsymbol{\hat s}$ or $-\boldsymbol{\hat s}$ depending on the funneling direction of the signal streamlines. Given the self-similar geometry of the signal field, we assume that the limit cycle chosen by the dynamics falls in the hemisphere including the prey direction (i.e., we choose the sign of $\boldsymbol{\hat e_1}$ such that $\boldsymbol{\hat e_1}\cdot \boldsymbol{r} <0$). 

\subsection{Performance metric}
To evaluate the performance of the  steer'n roll strategy,  we introduce the metric
	\begin{equation}\label{eq:P_definition}
		P(\hat{\ve{r}}) = -\langle \,\boldsymbol{\hat r}\, \cdot \boldsymbol{\hat t} \,\rangle\:,
	\end{equation}
which measures the mean alignment between the predator swimming direction $\boldsymbol{\hat t}$ and  the direction toward the prey given by $-\boldsymbol{\hat r} = -\boldsymbol{r}/\|\boldsymbol{r}\|$.
	The performance $P$ ranges from $-1$ to $1$. The value $P=1$ indicates a perfect alignment between $-\boldsymbol{\hat r}$ and $\boldsymbol{\hat t}$, $P=0$ indicates a random orientation, and $P=-1$ indicates anti-alignment.  
    
We calculate $P$ numerically at each position $\boldsymbol{r}$ by setting $V=0$ to isolate the rotational dynamics in \eqref{Eq:Dynamics}. The operator $\langle \cdot \rangle$ represents 
an ensemble average, performed by running $N = 10^3$ simulations with random initial orientations. The performance $P$ is evaluated in each simulation for $t =  10^3 / \omega_\mathrm{steer}$ to ensure that the dynamics has reached its limit cycle and to avoid the transient. 
    
    Because the prey-generated flow field is axisymmetric, 
    any meridional $(\rho, z)$--plane in cylindrical coordinates is equivalent to the $(x, z)$--plane. In addition,  the steer'n roll strategy is scale-free, so 
    the performance $P$ depends only on the predator’s angular position  relative to the prey, $\theta = \arctan(\rho/z)$.  

\subsection{Performance of steer'n roll}
\begin{figure*}[t]
		\centering
		\def\svgwidth{\textwidth}
		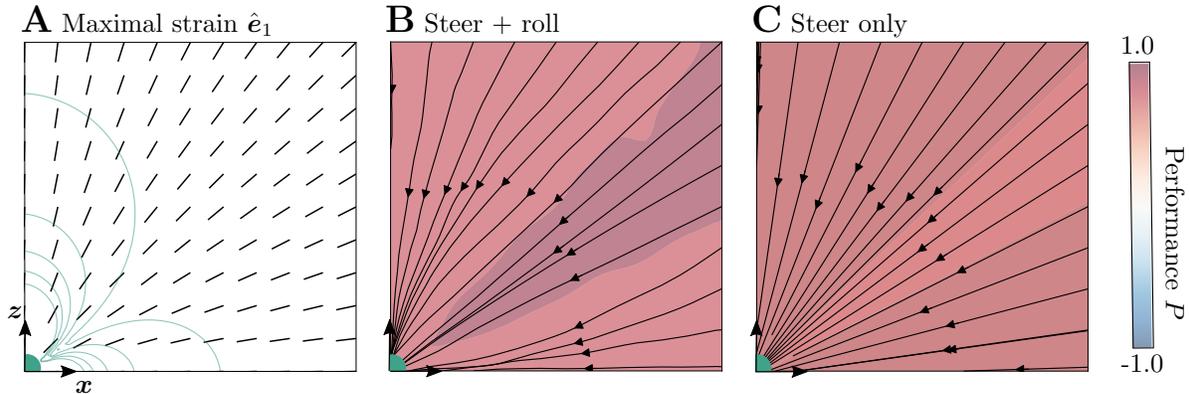
		\caption[Performance]{
\textbf{Performance of steer'n roll.}
The performance $P$ of the steer'n roll strategy for a predator detecting a stresslet flow field. 
(\textbf{A}) Direction of maximal strain $\boldsymbol{\hat{e}}_1$. The light green contour lines show the stresslet flow intensity. 
(\textbf{B}) Contour plot of the performance $P$ and average swimming direction $\boldsymbol{\hat{t}}$ for numerically integrated trajectories ($V=0$ and $\omega_\mathrm{roll} = 0.2\:\omega_\mathrm{steer}$). 
(\textbf{C}). Theoretical prediction of the performance $P$, and the mean predator's orientation $\braket{\boldsymbol{\hat{t}}} = \boldsymbol{\hat{e}}_1$, predicted from the fixed point analysis given in~\eqref{eq:fp}. There is no lengthscale in the plots because the problem is scale-free.}
    \label{fig:fig_3_performancetheory}
	\end{figure*}

Figure~\ref{fig:fig_3_performancetheory}B shows the performance $P$ for a stresslet flow generated by the prey. Superimposed streamlines indicate the long-time average orientation of the predator's orientation  $\langle \,\boldsymbol{\hat t} \,\rangle$. The absence of regions with $P \leq 0$ demonstrates that steer'n roll has a success rate of 100\%. 
The same results for the stokeslet flow field are provided in Fig.~\ref{fig:SokesletPerformance} in \textit{Supporting Information}. 

The predator's performance can also be evaluated analytically from \eqref{eq:fp}. According to this analysis, the limit cycle is centered around the eigen vector $\boldsymbol{\hat e_1}$ (Fig.~\ref{fig:fig_3_performancetheory}A), such that the predator's orientation is  $\langle\,\boldsymbol{\hat t}\,\rangle = \boldsymbol{\hat e_1}$ (Fig.~\ref{fig:fig_3_performancetheory}C).
The performance $P$ can also be evaluated analytically by assuming that $\ve{\hat{t}}$ samples uniformly the limit cycle (Fig.~\ref{fig:fig_3_performancetheory}C). The small discrepancy in the performance $P$ between the numerical solution in Fig.~\ref{fig:fig_3_performancetheory}B and the theoretical prediction in Fig.~\ref{fig:fig_3_performancetheory}C  likely originates 
from the uniform-sampling assumption of $\ve{\hat{t}}$ along the limit cycle in Fig.~\ref{fig:fig_3_performancetheory}C (see \textit{Supporting Information} Fig.~\ref{fig:DynamicDistr}). 

The good agreement between numerics and analytical predictions demonstrate that the analytical approach in Eqs.~(\ref{eq:fixedpoint}--\ref{eq:fp}) captures the mechanism by which stereoscopic flow sensing and the steer'n roll strategy enable the predator to localize the prey. 

\section{Robustness to Noise}

The sensorimotor modalities underlying predation are inherently noisy. We identify three sources of noise: (1) the predator may sense a noisy version of the flow signal generated by the prey; (2) its desired orientational response may be altered by a noisy execution; and (3) the background turbulent flow may add noise to the meaningful prey-generated flow signal. We assess the robustness of the steer'n roll strategy against each of these sources of noise.  
	
In the absence of noise, the steer'n roll strategy is scale-free: it does not depend on the prey-predator separation distance. However, the presence of noise introduces a characteristic length scale beyond which reliable localization breaks down. The goal of this section is to determine how length scale depends on the problem parameters for each possible source of noise.

\subsection{Sensory noise}\label{sec:TheoryMeasureNoise}
We first assess how sensory noise affects the predator's ability to infer the target direction. As shown in \eqref{eq:triangulation_d}, computing the target direction $\boldsymbol{\hat d}$ involves two steps:  determining the unit vector $\boldsymbol{\hat s}$  and evaluating the sign of the scalar $\xi$, given by \eqref{eq:triangulation_sign}. 
Our analysis shows that adding Gaussian noise to the signal field in \eqref{eq:dimensional_signal} has a comparatively small impact on the estimation of $\boldsymbol{\hat s}$, but a strong impact on the determination of the sign of $\xi$. We therefore focus on how noise influences $\xi$. 

For a prey generating a stresslet flow $\ve v_{\rm 2}$, given by \eqref{eq:Stresslet}, the leading order scales as $v_{\rm 2} \sim  a F / (8 \pi \mu r^2)$.
Let the velocity sensor of the predator be subject to a Gaussian noise of zero mean and standard deviation $v_\mathrm{noise}$. This noise affects the signal $\boldsymbol{S}$, given by \eqref{eq:dimensional_signal}, which scales as $ S \,\propto\, \ell v_{\rm 2} / r =  a F \ell / (8 \pi \mu r^3)$. Finally, from \eqref{eq:triangulation_sign}, we see that the sign of $\xi$ will likely  change when the noise is comparable to $\Delta \boldsymbol{S}= \boldsymbol{S^+} - \boldsymbol{S^-}$, given by \eqref{eq:stereosignal}, which scales as $\Delta S \propto 2\ell S / r =  a F \ell^2 / (4 \pi \mu r^4)$. By equating the scaling of $\Delta S$ with the scaling of the noise $
v_\mathrm{noise}$, we define a sensory distance:
    \begin{equation}
    r_\mathrm{sens} = \left(\frac{a \,\ell^2 F}{4\pi \mu v_\mathrm{noise}}\right)^{1/4}.
    \label{eq:r_sensory}
    \end{equation}
This distance is calculated for a stresslet flow field but it is easily generalizable to any singular solution of the Stokes equation.
When $r \ll r_\mathrm{sens}$, we predict that the steer'n roll strategy will be efficient, and when $r \gg r_\mathrm{sens}$, we predict random swimming as the predator is unable to determine the sign of $\xi$ because of noise. 
    
In Figure \ref{fig:fig_4_performancenoise}A, we compare these predictions  with numerical simulations in which Gaussian noise is added directly to the measured signals $\boldsymbol{S^+}$ and $ \boldsymbol{S^-}$. These results show that the performance metric $P$ remains unchanged for $r\lesssim r_\mathrm{sens}$, degrades around $r\sim r_\mathrm{sens}$, and becomes random for $r\gtrsim r_\mathrm{sens}$. This behavior confirms the scaling predicted in \eqref{eq:r_sensory}. 
	
\subsection{Strategy noise}
	To model strategy noise, we add rotational Brownian noise  to the predator's orientational dynamics in~\eqref{Eq:Dynamics}
	\begin{equation}\label{Eq:DynamicsNoise}
		\frac{d \boldsymbol{\hat e}}{dt}  = \boldsymbol{\omega} \times \boldsymbol{\hat e} +
		\omega_\mathrm{noise} \,\boldsymbol{\xi}  \times  \boldsymbol{\hat e} \:,
	\end{equation}
	where $\boldsymbol{\xi}(t)$ is a standard three-dimensional Wiener process and $\omega_\mathrm{noise}$ sets the noise strength, with units of angular velocity.  
    
    Figure~\ref{fig:fig_4_performancenoise}B shows how this noise-induced torque affects the performance $P$ and the predator's mean trajectories. As expected, orientational noise has little effect on performance as long as  $\omega_\mathrm{noise} \ll \omega_\mathrm{steer}$. This is because the noise is then insufficient to destroy the limit cycle in Fig.~\ref{fig:fig_3_t_trajectories} as a global attractor of the rotational dynamics; strategy noise simply scatters $\boldsymbol{\hat t}$ around the limit cycle without changing its mean.
    	
\subsection{Turbulence}	\label{sec:turbulentNoise}
Lastly, we consider the effect of background turbulence. We model  turbulence as a flow gradient that is uniform at the scale of the planktonic organisms.  For simplicity, we assume that this background flow is a pure strain with elongation along the $z$-axis and isotropic compression along the $x$ and $y$ axes. The resulting background flow can then be written as $\ve v(\ve x) = \dot{\gamma}_\mathrm{noise}\mathbb{E} \cdot \ve x$ where
	\begin{equation}
		\mathbb{E} = \frac{1}{\sqrt{6}}\begin{bmatrix}
			1 & 0 & 0 \\
			0 & 1 & 0 \\
			0 & 0 & -2
		\end{bmatrix},
	\end{equation}
	such that $\|\mathbb{E} \| = 1$. 
	
In the presence of this turbulence-induced background flow, the meaningful flow signal $\boldsymbol{S}$ measured by the predator is replaced by $\boldsymbol{S}  + \ell \dot{\gamma}_\mathrm{noise} \mathbb{E} \cdot \boldsymbol{\hat n}$. For a prey generating a stresslet,  the magnitude of the signal $\boldsymbol{S}$ scales as $S\,\sim\, \ell v_{\rm 2} / r =  a F \ell / (8 \pi \mu r^3)$. The meaningful prey-generated signal $S$ and the turbulent-induced contribution $\ell \dot{\gamma}_\mathrm{noise}$ become comparable when $r\sim r_\mathrm{turb}$, with
\begin{equation}\label{eq:r_turbulent}
    r_\mathrm{turb} = \left(
        \frac{a F}{8\pi\mu  \dot{\gamma}_\mathrm{noise}}
    \right)^{1/3}.     
\end{equation}    

\begin{figure*}[t]
		\centering
		\def\svgwidth{\textwidth}
		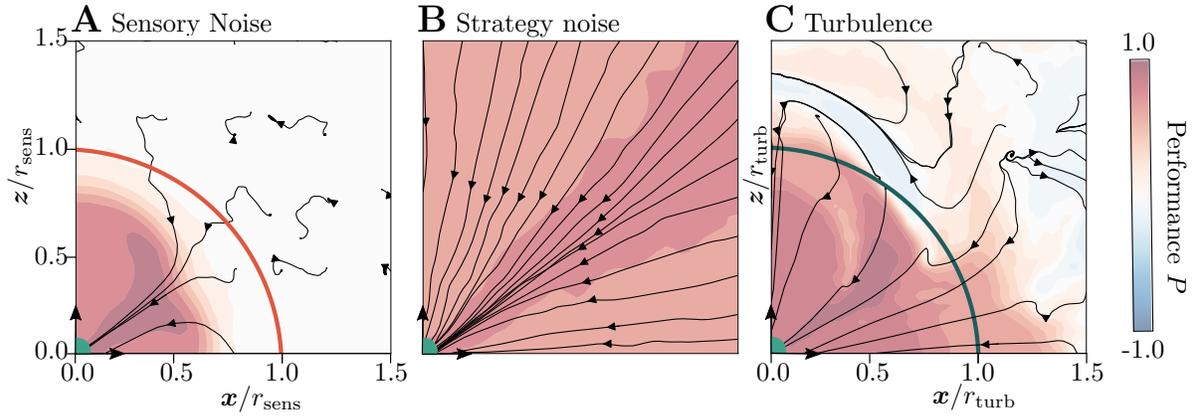
		\caption[Performance with noise]{
\textbf{Robustness to noise.} 
(\textbf{A}) Performance metric $P$ and average trajectories when flow-sensing is noisy. The red line correspond to the distance $r=r_\mathrm{sens}$, as given by Eq.~(\ref{eq:r_sensory}).  
(\textbf{B}) Performance $P$ when the predator’s rotational dynamics is affected by noise with $\omega_\mathrm{noise} = 0.5\, \omega_\mathrm{steer}$ (this noise does not introduce a scale in the problem and the strategy remains scale-free). 
(\textbf{C}) Performance $P$ in presence of a background flow modeling turbulence. The blue line represent the distance  $r=r_\mathrm{turb}$, as given by Eq.~(\ref{eq:r_turbulent}). 
All simulations are performed with $V = 0$ and $\omega_\mathrm{roll} = 0.2\,\omega_\mathrm{steer}$, similarly to Fig.~\ref{fig:fig_3_performancetheory}B.
}
	\label{fig:fig_4_performancenoise}
\end{figure*}

Figure~\ref{fig:fig_4_performancenoise}C compares this theoretical prediction with numerical simulations in which we added turbulent noise directly to the measured signals $\ve S^\pm$. When the prey-generated signal dominates the background noise, that is, for distances $r\lesssim r_\mathrm{turb}$, the steer’n roll strategy reliably localizes the prey. Beyond this region, localization performance degrades; however, localized regions with ``false positives'' are present. 
	
\section{Discussion}
We introduced steer’n roll, a bioinspired sensing and control strategy that enables the localization of a flow source in the Stokes regime by exploiting two flow measurements to resolve the fundamental ambiguity of Stokes flows~\cite{takagi2017}. Namely, by sampling the signal field at two spatially separated locations, a predator can triangulate the local signal directions and infer a swimming direction aligned with the funneling of the flow streamlines. This inferred direction breaks the symmetry inherent to Stokes-flow singularities and provides a robust cue for steering toward the source. We further examined how roll motions, commonly observed in microorganisms~\cite{Niimoto2020,Fenchel2001,wan2021origins, Cortese2021, Leptos2023}, enable predators with limited pitch control to escape planar motion and achieve effective three-dimensional reorientation. In the absence of noise, the strategy is scale-free, which makes it independent of the absolute spatial scale of the predator–prey system. 

This minimal sensing strategy is motivated by the biology of copepods, which possess arrays of highly sensitive mechanoreceptors distributed along their antennules~\cite{Strickler1973b, Yen2015}. In reality, these mechanosensory arrays provide a spatially extended sampling of the local flow field~\cite{Fields2002, Yen2015}. Here, we deliberately approximated this complex sensory apparatus by two spatially separated measurements of the flow gradient, capturing the minimal geometric information required to infer a direction toward the flow source. Within this approximation, steer'n roll provides a mechanistic interpretation of how distributed flow sensing can be exploited to extract directional information. 

Within the scope of this minimal model, we went beyond demonstrating a prey-detection mechanism, and we derived analytical expressions that quantify the limits of detection in the presence of multiple sources of noise: sensory noise, motor control noise, or environmental turbulence. Specifically, we provided expressions that relate the maximum detection distance to the predator inter-sensor distance $\ell$, the prey size $a$, and swimming speed $U$, and the intensity of sensory or environmental noise. We identified a critical distance at which the signal-to-noise ratio approaches unity, beyond which prey localization is not effective. 

\begin{figure}[th]
		\centering
		\def\svgwidth{\columnwidth}
        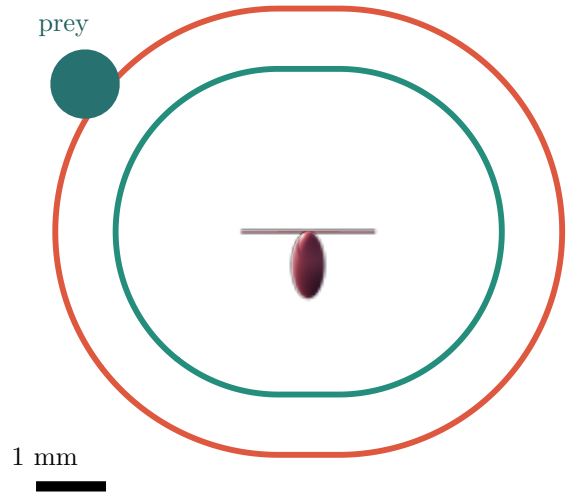
		\caption[Copepods Sensitivity]{The measured $r_{\rm sens}$ and $r_{\rm turb}$ distances from the antennae at which a prey (panel (\textbf{A})) or a mate (panel (\textbf{B})) can be sensed by the copepod. In panel (\textbf{A}), the gray dashed line show the average experimentally observed distance of copepods successful capture (data from~\cite{Jonnson1990}).}
		\label{fig:RealSituation}
	\end{figure}

As a concrete example, we consider the case of a predator copepod, \textit{Acartia tonsa},  with antennules of length $\ell=1\,$mm and sensory sensitivity of $v_{\mathrm{noise}} = 10\,\mu$m\,s$^{-1}$~\cite{Yen1992, Kiorboe1999}, placed in a turbulent flow with a strain rate $\dot{\gamma}_{\mathrm{noise}} = 0.1\,$s$^{-1}$, typical of open ocean turbulence~\cite{Fuchs2016}. 
We further consider two swimming prey subject to a drag force $F = 6 \pi \mu a U$: a unicellular ciliate, \textit{Strombidium reticulatum}, with $a = 2\times10^{-5}\,$m and $U = 10^{-3}\,$m\,s$^{-1}$ \cite{Jonnson1990}, and another copepod with $a=0.5\times 10^{-3}\,$m and $U=10^{-2}$m$\,$s$^{-1}$~\cite{Yen1998}.
The predicted detection distances, from \eqref{eq:r_sensory} and \eqref{eq:r_turbulent}, are $r_{\mathrm{sens}} \approx 0.49\,$mm and $r_{\mathrm{turb}} \approx 0.14\,$mm for the ciliate and $r_{\mathrm{sens}} \approx 5.0\,$mm and $r_{\mathrm{turb}} \approx 2.7\,$mm for the copepod (Fig.~\ref{fig:RealSituation}).  
The predictions of the maximal detection distance $r_{\mathrm{sens}}$ are in excellent agreement with experimental measurements on \textit{Strombidium reticulatum} \cite{Jonnson1990} as shown in Fig.~\ref{fig:RealSituation}A. We can further note that  $r_{\mathrm{sens}} > r_{\mathrm{turb}}$ for prey smaller than a few millimeters, showing that sensitivity is rarely the limiting factor for prey capture in typical turbulent environments. 

The present framework necessarily involves simplifying assumptions. In particular, the multipole expansion of the flow implicitly assumes that the prey is small compared to the predator-prey distance ($a\ll r$), and the two-sensor approximation assumes that the inter-sensor distance is small compared to the predator-prey distance ($\ell \ll r$). Small prey, such as the ciliate considered in Fig.~\ref{fig:RealSituation}A, challenge these assumptions. Moreover, the ecological dynamics of hunting or mating are more intricate than the idealized scenario considered here: the prey is also swimming, and both the prey and the predator are affected by each other's flows \cite{Borra2022}. 

Real copepods integrate multiple sensory modalities, including chemical cues and light sensitivity~\cite{Visser2011}, and the measured flow field itself contains additional information, such as pressure signals, that may contribute to prey localization~\cite{LegierVisser1986}. In natural environments, predators are also exposed to the superposition of flow disturbances generated by multiple sources, which may need to be disentangled to locate prey, predators, or mates~\cite{Jiang2008}. By isolating mechanosensation in a minimal setting, our work clarifies its distinctive role while providing a foundation for incorporating these additional layers of complexity.

Finally, our analysis focuses on instantaneous, spatial sensing of flow gradients. In realistic predator–prey or mating interactions, both agents move, and flow signals evolve over time along the predator’s trajectory. In such dynamic contexts, memory and temporal integration may play an important role in resolving ambiguities and improving localization performance. Extending the steer’n roll framework to incorporate memory would allow predators to exploit correlations in the sensed flow over time and may enhance performance in complex, time-dependent environments \cite{verano2023olfactory}. Exploring these effects is a natural direction for future work on flow-mediated interactions in planktonic systems.

\section{Acknowledgments}
This project has received funding from the European Research Council (ERC) under the European Union’s Horizon 2020 research and innovation program (grant agreement No 834238), from the National Science Foundation (NSF) through grants RAISE IOS-2034043 and CBET-2100209, and from the Office of Naval Research (ONR) through grants N00014-22-1-2655 and N00014-19-1-2035.

\noindent\textbf{Declaration of interests:} The authors declare no competing interests.
\bigskip
\bibliographystyle{apsrev}
\bibliography{references}
\vfill
~
\vfill
\newpage

%
%

\setcounter{section}{0}
\setcounter{page}{1}
\renewcommand{\thefigure}{S\arabic{figure}}
\setcounter{figure}{0}
\setcounter{equation}{0}
\renewcommand{\theequation}{S\arabic{equation}}

\begin{center}
    {\Large \textbf{Supporting Information }}
\end{center}


\section{Triangulation Algorithm}
\label{sec:AppTriang}

\begin{figure}[th]
    \centering
    \def\svgwidth{0.25\textwidth}
    \includegraphics[width=0.3\textwidth]{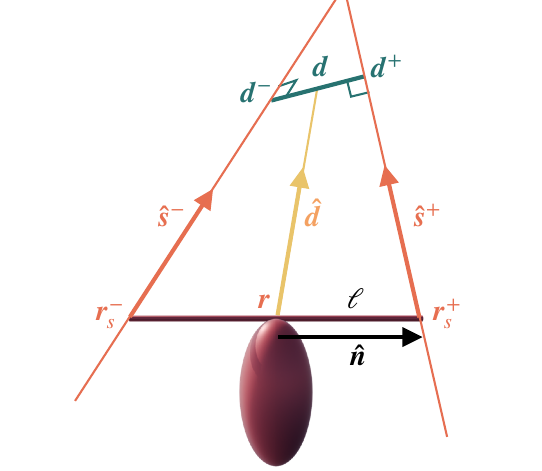}
    \caption[Triangulation]{Schematic of the vectorial triangulation operation. The target location $\boldsymbol{\hat{d}}$ is inferred from the two sensed signal vector $\hat{\ve s}^\pm$ originating from the sensor positions $\ve r^\pm$.}
		\label{fig:Triangulation}
	\end{figure}

Here, we provide a detailed derivation of the triangulation algorithm used to calculate the target direction $\boldsymbol{\hat{d}}$. 
The left and right sensors are located at the points $\boldsymbol{r}_s^\pm =\boldsymbol{r} \pm \ell \boldsymbol{\hat n}$. From each of these sensors passes a line in the direction $\boldsymbol{\hat{s}}^\pm$ (Fig.~\ref{fig:Triangulation}). Our goal is to determine the points $\boldsymbol{d}^\pm = \boldsymbol{r}_s^\pm + d^\pm \boldsymbol{\hat{s}}^\pm $ along these lines, such that the segment $\boldsymbol{d}^+ - \boldsymbol{d}^-$ corresponds to the minimum distance between the two lines. The lengths $d^\pm$ can be determined from the orthogonality between this segment and the two lines. Writing  $(\boldsymbol{d}^+ - \boldsymbol{d}^-)\cdot \boldsymbol{\hat{s}}^\pm = 0$ yields
\begin{equation}
    d^\pm = d \mp \frac{(\boldsymbol{r}_s^+ - \boldsymbol{r}_s^-)\cdot \boldsymbol{\hat{s}}^\mp}{1 + \boldsymbol{\hat{s}}^+\cdot \boldsymbol{\hat{s}}^-}, \quad \mbox{with }
    d = \frac{(\boldsymbol{r}_s^+ - \boldsymbol{r}_s^-)\cdot (\boldsymbol{\hat{s}}^+ - \boldsymbol{\hat{s}}^-)}{(\boldsymbol{\hat{s}}^+\cdot \boldsymbol{\hat{s}}^-)^2 - 1}.
\end{equation}

The segment joining the midpoint $\boldsymbol{r} = (\boldsymbol{r}^+ + \boldsymbol{r}^-)/2$ between the sensors and the midpoint $\boldsymbol{d} = (\boldsymbol{d}^+ + \boldsymbol{d}^-)/2$ between $\boldsymbol{d}^+$ and $\boldsymbol{d}^-$ is given by
\begin{equation}
    \boldsymbol{d} - \boldsymbol{r} = d \frac{\boldsymbol{\hat{s}}^+ + \boldsymbol{\hat{s}}^-}{2},
\end{equation}
such that the funneling direction is 
\begin{equation}\label{eq:d_exact}
    \boldsymbol{\hat{d}} = \frac{\boldsymbol{d} - \boldsymbol{r}}{\| \boldsymbol{d} - \boldsymbol{r}\|}= \mathrm{sign}(\xi)\frac{\boldsymbol{\hat{s}}^+ + \boldsymbol{\hat{s}}^-}{\|\boldsymbol{\hat{s}}^+ + \boldsymbol{\hat{s}}^-\|},
\end{equation}
where $\xi = -\boldsymbol{\hat{n}}\cdot (\boldsymbol{\hat{s}}^+ - \boldsymbol{\hat{s}}^-)/2\ell$ is a scalar with the same sign as $d$. 
Note that the above derivation has been obtained without any hypothesis. In the far field, when $\ell \ll r$, \eqref{eq:d_exact} can be simplified into \eqref{eq:triangulation_d} given in the main text.

\section{Analysis of Rotational Dynamics}
\label{sec:TriangulationTheoretical}

Here, we provide the complete mathematical derivation of the fixed points of the orientation dynamics.  
As said in the main text, we  consider $V=0$ and $\omega_\mathrm{roll} = 0$  and look for the solution $\boldsymbol{\hat t^*}$ and $\boldsymbol{\hat n^*}$ of 
\begin{equation}\label{eq:fixed_appendix}
    \boldsymbol{\hat t}^* \times \left(
            \boldsymbol{\hat n}^* \cdot \nabla \ve v
        \right)
		= 0,
\end{equation}
with the flow gradient expressed as
\begin{equation}\label{eq:gradient_suppl}
		\nabla \boldsymbol{v} = \lambda_1
		\begin{bmatrix}
			1 		& \Omega 	& 0 \\
			-\Omega & \tilde{\lambda} 		& 0 \\
			0 		& 0 		& -1-\tilde{\lambda}
		\end{bmatrix},
	\end{equation}
in the basis of the eigenvectors $(\hat{\ve e}_1, \hat{\ve e}_2, \hat{\ve e}_{3})$ of its symmetric part. Here, $\lambda_1$ is the largest eigenvalue in absolute value and $\hat{\ve e}_1$ the corresponding eigenvector. The ratio
$\tilde{\lambda} = \lambda_2 / \lambda_1$ is the ratio between the eigenvalues in the meridional plane, with $-1\leq \tilde{\lambda} \leq 1$.

The first two components of \eqref{eq:fixed_appendix} can be combined to find two relations between the components $n^*_i$ and $t^*_i$ of $\boldsymbol{\hat n^*}$ and $\boldsymbol{\hat t^*}$ in the $(\hat{\ve e}_1, \hat{\ve e}_2, \hat{\ve e}_{3})$ basis  
\begin{equation}
\begin{split}
n^*_1 & = - \frac{n^*_3}{t^*_3} \frac{1 + \tilde{\lambda}}{\Omega^2 + \tilde{\lambda}} \left(\tilde{\lambda} t^*_1 + \Omega t^*_2 \right) ,\\
n^*_2 & = - \frac{n^*_3}{t^*_3} \frac{1 + \tilde{\lambda}}{\Omega^2 + \tilde{\lambda}} \left(- \Omega t^*_1 + t^*_2 \right).
\end{split}
\end{equation}
Inserting these expressions into the equation $\boldsymbol{\hat n^*}\cdot \boldsymbol{\hat t^*} = 0$ gives
\begin{equation}
    -\frac{1 + \tilde{\lambda}}{\Omega^2 + \tilde{\lambda}} \left(\tilde{\lambda} t^{*2}_1 + t^{*2}_2 \right) + t^{*2}_3 =0, 
\end{equation}
which, after exploiting the unit norm of $\boldsymbol{\hat t^*}$, gives Eq.~(\ref{eq:fp}a) relating $t^*_2$ and $t^*_3$ found in the main text. 

\begin{figure*}[h]
	\centering
\def\svgwidth{\textwidth}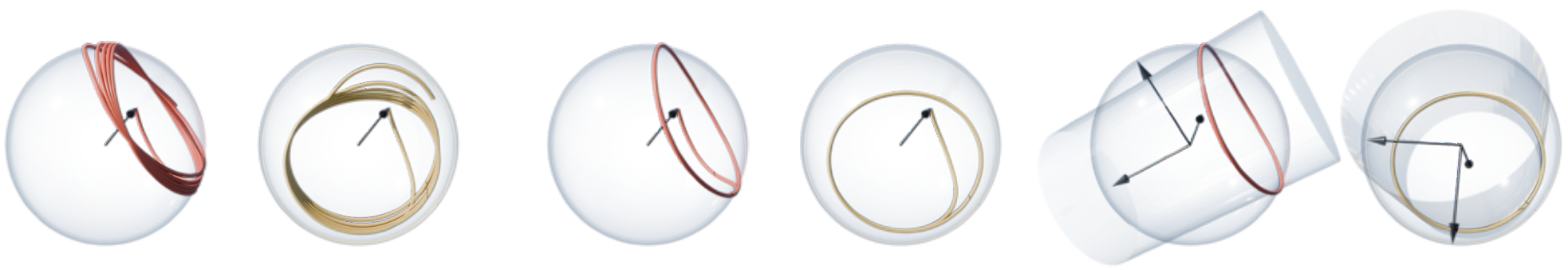
	\caption[n trajectories]{
    \textbf{Predator's orientation dynamics.}
(\textbf{A}) Evolution of \(\boldsymbol{\hat n}\) extracted from the trajectories shown in Fig.~\ref{fig:fig_2_trajectories}B. The dynamics converges to an imperfect limit cycle.
(\textbf{B}) Same as A, but in the absence of rotation (\(V=0\)), showing convergence to a perfect limit cycle.
(\textbf{C}) Theoretical limit cycle predicted for $V=0$ and $\omega_\mathrm{roll}=0$, by the fixed-point analysis (see~\eqref{eq:fp}).}
	\label{fig:Trajectories_vector_n}
	\end{figure*}

A similar calculation leads to Eq.~(\ref{eq:fp}b) relating $n^*_2$ and $n^*_3$, which also correspond to an elliptic cylinder with axis $\hat{\ve e}_1$ (Fig.~\ref{fig:Trajectories_vector_n}).

\section{Distribution of the Swimming Direction over the Limit Cycle}
\label{app:Distribution}

From the theoretical analysis of the predator's orientational dynamics, we know that the swimming direction $\hat{\ve t}$ converges towards the limit cycle $\hat{\ve t}^*$, independently of the initial orientation. This limit cycle lies at the intersection of an elliptic cylinder of axis $\hat{\ve e}_1$ and the unit sphere. We will now calculate the performance $P$ when the limit cycle is sampled uniformly.

To evaluate the performance $P$, we calculate the scalar product $-\hat{\ve r} \cdot \hat{\ve t}^*$ in the basis $(\hat{\ve e}_1, \hat{\ve e}_2, \hat{\ve e}_3)$, where $\hat{\ve t}^*$ is given by Eq.~(\ref{eq:fp}a)
\begin{equation}
    \hat{\ve t} = \left[\sqrt{1 - a_2^2\cos^2\alpha - a_3^2\sin^2\alpha}, a_2 \cos\alpha, a_3 \sin\alpha \right],
\end{equation}
with $a_2 = \sqrt{\tilde{\lambda}/(\tilde{\lambda}-1)}$, and $a_3 = \sqrt{(\tilde{\lambda}^2 +\tilde{\lambda})/(\tilde{\lambda}^2+ 2\tilde{\lambda}+\Omega^2)}$. 
Then averaging over $\alpha$ yields
\begin{equation}\label{eq:Panalytical}
\begin{split}    
    P & = - \frac{1}{2 \pi} \int_0^{2 \pi} r_1 \sqrt{1 - a_2^2\cos^2\alpha - a_3^2\sin^2\alpha}\, d\alpha, \\
    & = -\frac{2 r_1 \sqrt{1-a_2^2}}{\pi} E\left(\frac{a_2^2-a_3^2}{a_2^2-1}\right)\approx - r_1 \sqrt{1 - a_2a_3},
\end{split}
\end{equation}
where $E$ is the complete elliptic integral. 

\begin{figure*}[h!]
		\centering
\def\svgwidth{\textwidth}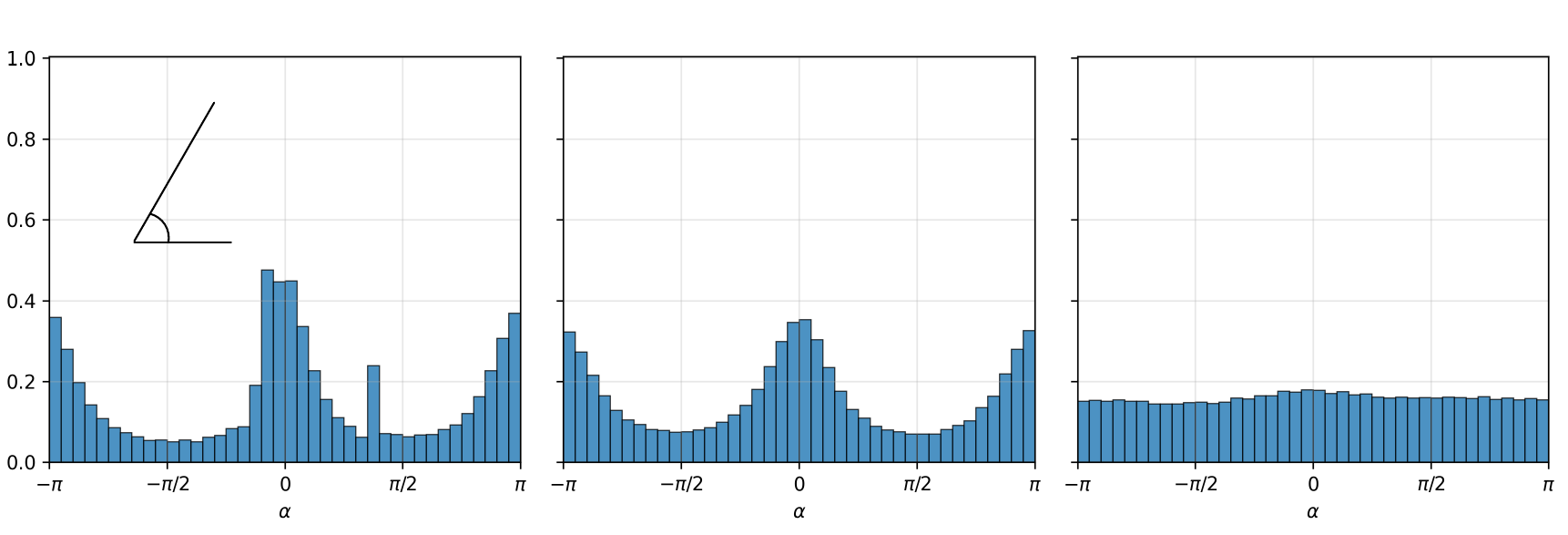
		\caption[Probability density function of $P$.]{
This figure shows the histograms of $\alpha$, the angular position along the limit cycle of the swimming vector $\boldsymbol{\hat{t}}$ after the convergence of the orientation dynamics, for distinct predator positions. The problem is scale free and the flow measured axisymmetric, so the dynamics depends only upon the angular position of the predator with respect to the flow symmetry axis $\boldsymbol{\hat{z}}$. The angle $\alpha$ is inferred as $\arctan( t_3a_2/t_2a_3)$. As shown, the distribution is not uniform. Simulations here are run for $V=0$ and $\omega_{\rm roll} =0.2\,\omega_{\rm steer}$.}
\label{fig:DynamicDistr}
	\end{figure*}

In Fig.~\ref{fig:fig_3_performancetheory}, there is a small discrepancy between the numerically computed performance $P$ and the value inferred from \eqref{eq:Panalytical}. There are two reasons for this discrepancy. First, the non zero roll in the numerical simulations slightly modify the limit cycle. Second, the limit cycle is not sampled uniformly along the elliptic cylinder as can be seen from Fig.~\ref{fig:DynamicDistr}. 

\section{Absence of roll for predators with limited yaw motions}
\label{app:LimitedYaw}

\begin{figure*}[h!]
	\centering
	\def\svgwidth{\textwidth}
	\includegraphics[width=\textwidth]{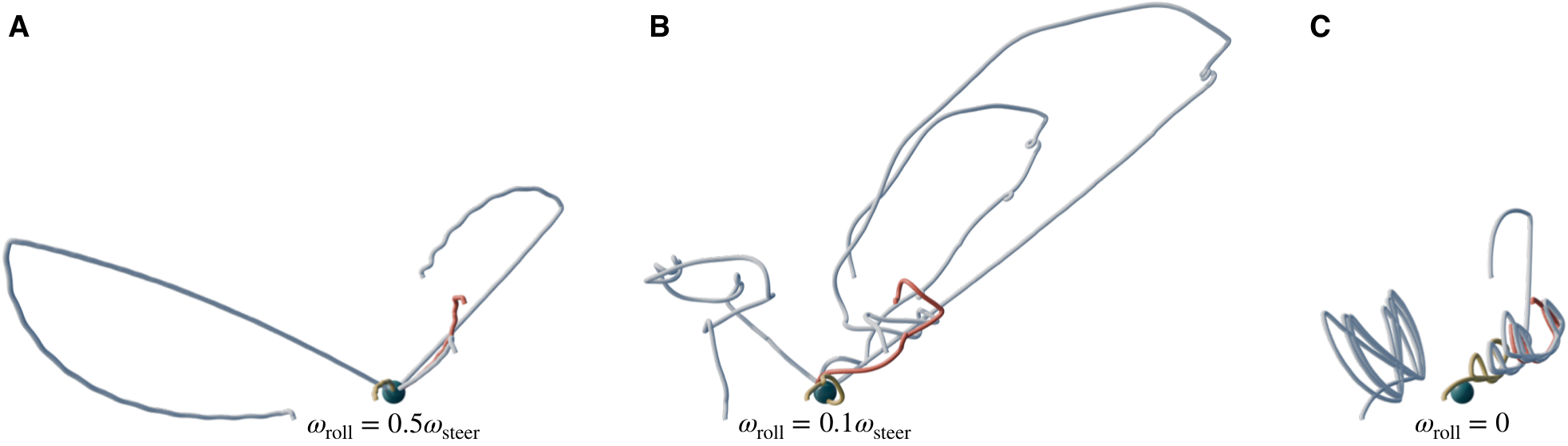}
	\caption[Scalar measure]{Same as Fig.~\ref{fig:fig_2_trajectories} for a predator limited to yaw motions (i.e., $\ve \omega_\mathrm{steer}$ projected onto $\ve{\hat b}$), for different values of $\omega_{\rm roll}$ as labelled. The steer'n roll strategy is unsuccessful when $\omega_{\rm roll}=0$, because the predator cannot escape its initial $(\ve{\hat t}, \ve{\hat n})$--plane.}
	\label{fig:RollScalar}
	\end{figure*}

A predator able to yaw only in one plane may face limitations in detecting the prey in absence of roll. As shown in Fig.~\ref{fig:RollScalar}, in absence of roll, the predator fails in detecting the prey.

\section{Stokeslet flow field}
\label{app:StokesletPerformance}

\begin{figure*}[ht!]
	\centering
\def\svgwidth{\textwidth}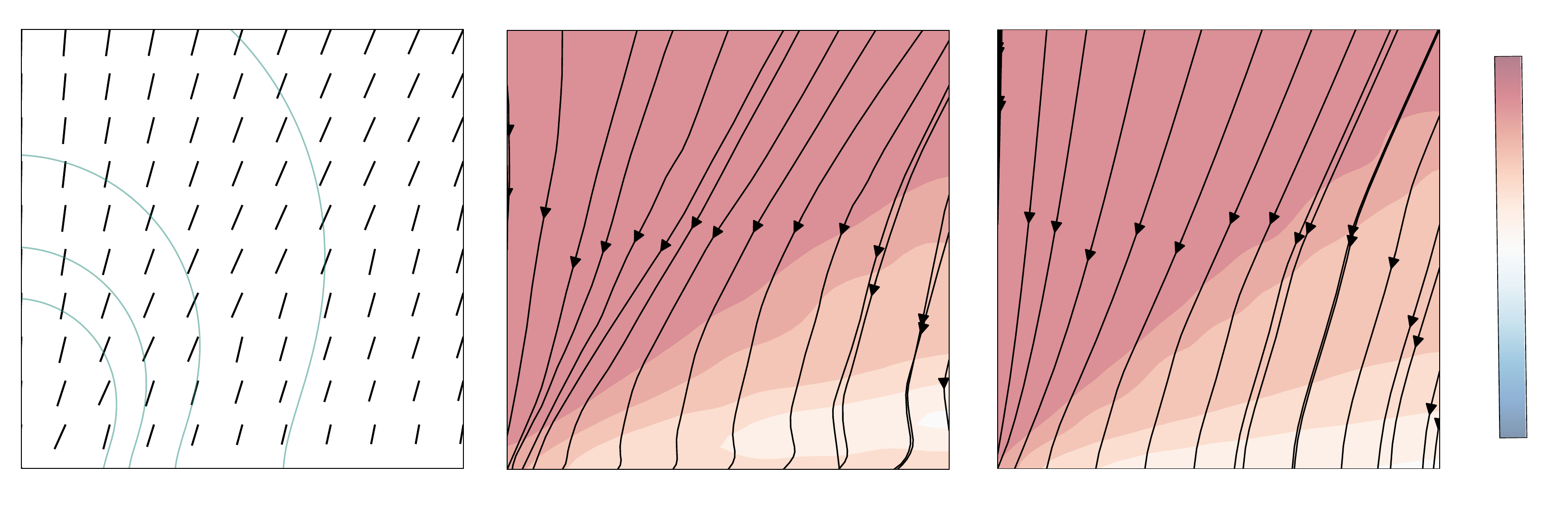
	\caption[Performance stokeslet]{
\textbf{Performance of steer'n roll.}
The performance $P$ of the steer'n roll strategy for a predator detecting a stokeslet flow field. 
(\textbf{A}) Direction of maximal strain $\boldsymbol{\hat{e}}_1$. The light green contour lines show the stresslet flow intensity. 
(\textbf{B}) Contour plot of the performance $P$ and average swimming direction $\boldsymbol{\hat{t}}$ for numerically integrated trajectories ($V=0$ and $\omega_\mathrm{roll} = 0.2\:\omega_\mathrm{steer}$). 
(\textbf{C}). Theoretical prediction of the performance $P$, and the mean predator's orientation $\braket{\boldsymbol{\hat{t}}} = \boldsymbol{\hat{e}}_1$, predicted from the fixed point analysis given in~\eqref{eq:fp}.}
	\label{fig:SokesletPerformance}
	\end{figure*}

The performance $P$ of the steer'n roll strategy can be analytically computed also for a stokeslet flow field. Interestingly, the limit cycle for the stokeslet gradient is a circle. One of the eigenvector of the strain matrix for this flow is degenerate. Therefore, the axis of the limit cicle ellipse are equivalent. Results are provided in Fig.~\ref{fig:SokesletPerformance}. The same calculation can be done for any Stokes singularity.

\end{document}